\documentclass[aps,eqsecnum,preprint,floats,epsf,epsfig,nofootinbib,showpacs]{revtex4}
\usepackage{graphicx}
\def\be{\begin{eqnarray}}
\def\en{\end{eqnarray}}
\def\non{\nonumber}
\def\la{\langle}
\def\ra{\rangle}

\def\prl{{ Phys. Rev. Lett.}~}

\def\bi{\bibitem}

\begin{document}

\title{\Large \bf $SU(3)$ symmetry breaking in decay constants and electromagnetic
properties of pseudoscalar heavy mesons
 }

\author{ \bf  Chien-Wen Hwang\footnote{
t2732@nknucc.nknu.edu.tw}}

\affiliation{\centerline{Department of Physics, National Kaohsiung Normal University,} \\
\centerline{Kaohsiung, Taiwan 824, Republic of China}
 }


\begin{abstract}
In this paper, the decay constants and mean square radii of
pseudoscalar heavy mesons are studied in the $SU(3)$ symmetry
breaking. Within the light-front framework, the ratios $f_{D_s}/f_D$
and $f_{B_s}/f_B$ are individually estimated using the hyperfine
splittings in the $D_{(s)}^*-D_{(s)}$ and $B_{(s)}^*-B_{(s)}$ states
and the light quark masses, $m_{s,q}$ ($q=u,d$), to extract the wave
function parameter $\beta$. The values $f_{D_s}/f_D= 1.29\pm0.07$
and $f_{B_s}/f_B= 1.32\pm 0.08$ are obtained, which are not only
chiefly determined by the ratio of light quark masses $m_s/m_q$, but
also insensitive to the heavy quark masses $m_{c,b}$ and the decay
constants $f_{D,B}$. The dependence of $f_{B_c}/f_B$ on $\Delta
M_{B_cB^*_c}$ with the varied charm quark masses is also shown. In
addition, the mean square radii are estimated as well. The values
$\sqrt{\langle r^2_{D_s^+}\rangle/\langle r^2_{D^+}\rangle}
=0.740^{-0.041}_{+0.050}$ and $\sqrt{\langle
r^2_{B_s^0}\rangle/\langle r^2_{B^0}\rangle}
=0.711^{-0.049}_{+0.058}$ are obtained, and the sensitivities of
$\langle r^2_P \rangle$ on the heavy and light quark masses are
similar to those of the decay constants.
\end{abstract}
\maketitle %
\section{Introduction}
The decay constants of pseudoscalar heavy mesons with $c$ and $b$
quarks play an important role for studies of CP violation and in
extracting the Cabibbo-Kobayashi-Maskawa matrix elements.
Experimentally, new data on the charm meson decay constants $f_D$
and $f_{D_s}$ have been reported \cite{PDG08,CLEO09}. As the
calculations of the decay constants are related to the wave function
overlap of the quark and antiquark which are governed by the strong
interaction, they therefore provide a crucial manner to compare
different theoretical methods. In addition, the determination of
$f_{B_S}$ remains beyond the reach of current experiments, thus the
reliability of estimated $f_{B_s}$ by a theoretical approach is
dependent on whether the determinations of $f_D$ and $f_{D_s}$ by
this approach are consistent with the new data. During the last
decade, the decay constants of pseudoscalar heavy mesons have been
studied in lattice simulations
\cite{HU,QCDSF,CHIU,MILC,HPQCD,UKQCD,Bec}, in the QCD sum rules
approach \cite{BPS,BPS1,Nar,Jamin}, and in the relativistic quark
model \cite{bada,EFG,BS2,BS1,Choi07}.

The understanding of the electromagnetic (EM) properties of hadrons
is also an important topic, and the EM form factors which are
calculated using nonperturbative methods are the useful tool for
this purpose. There have been numerous experimental
\cite{Bebek,Brauel,Amen1,Amen2,Dally,Volmer} and theoretical studies
\cite{RA} of the EM form factors of the light pseudoscalar meson
($\pi$ and $K$). However, the EM form factors of heavy mesons (which
contain one heavy quark) have much fewer studies
\cite{Braz,latticeQ} than those of light ones. The present paper is
devoted to an analysis of the wave function and decay constant by
the hyperfine mass splitting of heavy mesons and the formulas of the
decay constant and mean square radius within the light-front (LF)
framework. We present the $SU(3)$ symmetry breaking effect in decay
constants and electromagnetic properties of pseudoscalar heavy
mesons.

The light-front quark model (LFQM) is a promising analytic method
for solving the nonperturbative problems of hadron physics
\cite{BPP}, as well as offering many insights into the internal
structures of bound states. The basic ingredient in LFQM is the
relativistic hadron wave function which generalizes distribution
amplitudes by including transverse momentum distributions and
contains all the information of a hadron from its constituents. The
hadronic quantities are represented by the overlap of wave functions
and can be derived in principle. The light-front wave function is
manifestly a Lorentz invariant, expressed in terms of internal
momentum fraction variables which are independent of the total
hadron momentum. Moreover, the fully relativistic treatment of quark
spins and center-of-mass motion can be carried out using the
so-called Melosh rotation \cite{LFQM}. This treatment has been
successfully applied to calculate phenomenologically many important
meson decay constants and hadronic form factors \cite{Jaus1,Jaus91,
CCH1, Jaus2, Choi, CCH2, Hwang}.

The remainder of this paper is organized as follows. In Sec. II an
analysis of wave function and decay constant is presented. In Sec.
III the formulism of LFQM is reviewed briefly, and the formulae of
decay constant and mean square radius are derived. In Sec. IV
numerical results and discussions are presented. Finally, the
conclusions are given in Sec. V.

\section{Analyses of wave function and decay constant}
The decay constant $f_P$ for a pseudoscalar meson is defined by a
matrix element of the axial vector current between the vacuum and
the meson bound state:
 \be
 \langle 0 |\bar q_1 \gamma_\mu \gamma_5 q_2 | P(P)\rangle=if_P
 P_\mu. \label{definefp}
 \en
In a nonrelativistic approximation, $f_P$ is related to the
Bethe-Salpeter wave function at the origin $|\Psi(0)|$ as
\cite{PT,Kras,HW}
 \be
 f_P\simeq \frac{2\sqrt{N_c}}{\sqrt{M_P}} |\Psi(0)|,\label{NR}
 \en
where $N_c$ is the color number and $M_P$ is the mass of the meson.
If we consider the potential of a hyperfine interaction inside the
meson to $O(\alpha_s)$:
 \be
 V_{\rm hf}=\frac{4\alpha_s(3 \vec {s_1}\cdot \hat{r}
 \vec {s_2}\cdot \hat{r}-\vec {s_1} \cdot \vec {s_2})}{3 m_1 m_2 r^3}+
 \frac{32\pi\alpha_s {\vec {s_1} \cdot \vec{s_2}}}{9 m_1 m_2}
 \delta^3(\vec r), \label{hf}
 \en
where $\alpha_s$ is the strong coupling constant, $s_{1,2}(m_{1,2})$
are the spins (masses) of the constituent quark. For the $s$-wave
meson, the first term of Eq. (\ref{hf}) has no contribution and the
second term can distinguish the pseudoscalar and vector mesons.
Therefore, the hyperfine mass splitting is obtained as
 \be
 \Delta M_{PV} = \frac{32\pi\alpha_s}{9 m_1 m_2}
 |\Psi(0)|^2. \label{deltaM}
 \en
By combining Eqs. (\ref{NR}) and (\ref{deltaM}) and canceling
$|\Psi(0)|$, we obtain:
 \be
 f_P=\left(\frac{27 \Delta M_{PV} m_1 m_2}{8\pi \alpha_s
 M_P}\right)^{1/2}.
 \en
If we suppose the strong coupling constants $\alpha_s(D)\simeq
\alpha_s(D_s)$ and $\alpha_s(B)\simeq \alpha_s(B_s)$, then the
ratios of the decay constants can be obtained as
 \be
 \frac{f_{D_s}}{f_D}=\left(\frac{\Delta M_{D_sD^*_s}}{\Delta M_{DD^*}}
 \frac{M_D}{M_{D_s}}\frac{m_s}{m_q}\right)^{1/2},\quad
 \frac{f_{B_s}}{f_B}=\left(\frac{\Delta M_{B_sB^*_s}}{\Delta M_{BB^*}}
 \frac{M_B}{M_{B_s}}\frac{m_s}{m_q}\right)^{1/2},\label{ratiof}
 \en
where $q=u,d$. From Eq. (\ref{ratiof}), we find that the ratios are
dependent on the ratio of light quark masses $m_s/m_q$ and are
independent of heavy quark masses. Furthermore, by canceling the
ratio $m_s/m_q$, we have a relation which does not contain any
parameter in the nonrelativistic approximation:
 \be
 \frac{f_{B_s}}{f_B}=\left(\frac{\Delta M_{DD^*}\Delta M_{B_sB_s^*}}
 {\Delta M_{D_sD_s^*}\Delta M_{BB^*}}
 \frac{M_{D_s}M_B}{M_DM_{B_s}}\right)^{1/2}\frac{f_{D_s}}{f_D}.
 \label{noparameter}
 \en

If one wants to include the relativistic correction to the ratios of
the decay constants, not only the values of $m_{s,q}$, but also the
form of wave function $\Psi(\vec{r})$ must be known. Let us come
back to Eq. (\ref{NR}). The deviation of Eq. (\ref{NR}) uses the
Fourier transform
 \be
 \Psi(\vec{r})=\int \frac{d^3k}{(2\pi)^{3/2}} e^{i \vec{r}\cdot
 \vec{k}}\phi(\vec{k}),
 \en
where $\phi(\vec{k})$ is the wave function in the momentum space. If
the Fourier transform is evaluated for the positive-energy
projection of the Bethe-Salpeter wave function at equal ``time"
$z^+=z^0+z^3=0$, then \cite{LB}
 \be
 f_P \sim \int \frac{dx d^2 k_\bot}{2(2\pi)^3} \phi(x,k_\bot),
 \en
where $x$ is the longitudinal momentum fraction, $k_\bot$ are the
relative transverse momenta, and $\phi(x,k_\bot)$ satisfies the
normalization condition:
 \be\label{normal}
 \int \frac{dx d^2 k_\bot}{2(2\pi)^3} |\phi(x,k_\bot)|^2=1.
 \en
In general, the momentum distribution amplitude $\phi(x,k_\bot)$ is
obtained by solving the light-front QCD bound state equation $H_{LF}
|P\rangle = M|P\rangle$ which is the familiar Schr\"{o}dinger
equation in ordinary quantum mechanics, and $H_{LF}$ is the
light-front Hamiltonian. However, at the present time, how to solve
the bound state equation is still unknown. Alternatively, we come
back to the still-unknown wave function $\Psi (\vec {r})$ 
and express it as a linear combination of the arbitrary known
functions which form a complete set. Of course, the complete set is
not unique. Here we give two examples for comparison. One is the
solution of $1/r$ potential \cite{textbook}:
 \be
 \Psi_{nlm}^c(\vec{r})=\left(\frac{2 \beta}{n}\right)^{3/2}
 \left[\frac{(n-l-1)!}{2 n[(n+l)!]^3}\right]^{1/2}\left(\frac{2\beta r}{n}\right)^l
 L^{2 l+1}_{n-l-1}(2\beta r/n){\rm exp}
 \left(-\frac{\beta r}{n}\right)Y_{lm}, \label{isoL}
 \en
where superscript $c$ means ``Coulomb," $\beta$ is a parameter which
has the energy dimension, $Y_{lm}$ is the spherical harmonics, and
$L^p_{q-p}(x)$ is an associated Laguerre polynomial which is defined
as
 \be
 L^p_{q-p}(x) \equiv (-1)^p\left(\frac{d}{dx}\right)^p e^x
 \left(\frac{d}{dx}\right)^q (e^{-x}x^q).
 \en
The other complete set is the solution of an isotropic harmonic
oscillator \cite{textbook}:
 \be
 \Psi_{nlm}^g(\vec{r})=\frac{\beta^{3/2}}{\pi^{1/4}}(\beta r)^l
 h_{nl}(\beta r){\rm exp}
 \left(-\frac{\beta^2 r^2}{2}\right)Y_{lm},
 \label{isoG}
 \en
where superscript $g$ means Gaussian and the first few $h_{nl}(\beta
r)$'s are
 \be
 h_{00}=2,\quad
 h_{11}=\sqrt{\frac{8}{3}},\quad
 h_{22}=\frac{4}{\sqrt{15}},\quad
 h_{20}=\sqrt{6}\left(1-\frac{2}{3}\beta^2 r^2\right).
 \en
Then the exact solution can be expressed as
 \be
 \Psi (\vec
 {r})=\sum_{nlm}^{\infty}a^{c(g)}_{nlm}\Psi_{nlm}^{c(g)}(\vec{r}),\label{combine}
 \en
where $\sum_{nlm}^{\infty}|a^{c(g)}_{nlm}|^2=1$. This way seems very
clumsy because, apart from $\beta$, it also introduces a series of
undetermined coefficients, $a_{nlm}$. The following considerations,
however, improve the situation. First, only the coefficients
$a_{n00}$ survive because we just studied the $s$-wave meson.
Second, we substitute Eq. (\ref{combine}) to Eq. (\ref{deltaM}) and
obtain
 \be
 \Delta M_{PV} = \frac{32\pi\alpha_s}{9 m_1
 m_2}\left(\frac{\beta^3}{4\pi^{3/2}}\right)
 \left[\sum_{{\rm even}~n}^\infty a^g_{n00}
 h_{n0}(0)\right]^2,\label{reduce}
 \en
which takes the Gaussian case, for example. It is worth noting that
the square bracket in Eq. (\ref{reduce}) is independent of
$\beta$. 
Then, the ratio of hyperfine mass splittings can be reduced as:
 \be
 \frac{\Delta M_{D_sD_s^*}}{\Delta M_{DD^*}}=\frac{m_q}{m_s}
 \left(\frac{\beta_{cs}}{\beta_{cq}}\right)^3, \qquad
 \frac{\Delta M_{B_sB_s^*}}{\Delta M_{BB^*}}=\frac{m_q}{m_s}
 \left(\frac{\beta_{bs}}{\beta_{bq}}\right)^3, \label{splitting}
 \en
which does not include any coefficient $a_{nlm}$. Equation
(\ref{splitting}) is also suitable to the Coulomb case. In fact, due
to the $\delta$ function in Eq. (\ref{hf}) having the dimension of
an energy cube and $\vec r$ is vanishing here, Eq. (\ref{splitting})
holds for any wave function which contains only one hadronic
parameter $\beta$. In addition, the ratios $\beta_{cs}/\beta_{cq}$
and $\beta_{bs}/\beta_{bq}$ in Eq. (\ref{splitting}) are mainly
influenced by the ratio $m_s/m_q$. This situation leads to the
ratios $f_{D_s}/f_{D}$ and $ f_{B_s}/f_B$ are sensitive to the
$SU(3)$ symmetry breaking, but are insensitive to the heavy quark
masses, which will be shown later. In the literature, there are some
early attempts \cite{Kh1,Kh2} to account for flavor symmetry
breaking in pseudoscalar meson decay constants. In the next section
the wave function $\Psi(\vec r)$ and the values of $m_{s,q}$ are
studied within the light-front framework.

\section{Light-Front Framework }
\subsection{General Formulism}

An $s$-wave meson bound state, consisting of a quark $q_1$ and an
antiquark $\bar q_2$ with total momentum $P$ and spin $J$, can be
written as (see, for example \cite{CCH1})
 \be
 |M(P, S, S_z)\rangle =\int &&\{d^3p_1\}\{d^3p_2\} ~2(2\pi)^3
 \delta^3(\tilde P -\tilde p_1-\tilde p_2)~\non\\
 &&\times \sum_{\lambda_1,\lambda_2}
 \Phi^{SS_z}(\tilde p_1,\tilde p_2,\lambda_1,\lambda_2)~
 |q_1(p_1,\lambda_1) \bar q_2(p_2,\lambda_2)\rangle,\label{lfmbs}
 \en
where $p_1$ and $p_2$ are the on-mass-shell light-front momenta,
 \be
 \tilde p=(p^+, p_\bot)~, \quad p_\bot = (p^1, p^2)~,
 \quad p^- = \frac{m_q^2+p_\bot^2}{p^+},
 \en
and
 \be
 &&\{d^3p\} \equiv \frac{dp^+d^2p_\bot}{2(2\pi)^3}, \nonumber \\
 &&|q(p_1,\lambda_1)\bar q(p_2,\lambda_2)\rangle
 = b^\dagger(p_1,\lambda_1)d^\dagger(p_2,\lambda_2)|0\rangle,\\
 &&\{b(p',\lambda'),b^\dagger(p,\lambda)\} =
 \{d(p',\lambda'),d^\dagger(p,\lambda)\} =
 2(2\pi)^3~\delta^3(\tilde p'-\tilde p)~\delta_{\lambda'\lambda}.
 \nonumber
 \en
In terms of the light-front relative momentum variables $(x,
k_\bot)$ defined by
 \be
 && p^+_1=(1-x) P^{+}, \quad p^+_2=x P^{+}, \nonumber \\
 && p_{1\bot}=(1-x) P_\bot+k_\bot, \quad p_{2\bot}=x
 P_\bot-k_\bot,
 \en
the momentum-space wave-function $\Psi^{SS_z}$ can be expressed as
 \be
 \Phi^{SS_z}(\tilde p_1,\tilde p_2,\lambda_1,\lambda_2)
 = \frac{1}{\sqrt N_c}
 R^{SS_z}_{\lambda_1\lambda_2}(u,\kappa_\bot)~ \phi(x,
 k_\bot),\label{Psi}
 \en
where 
$R^{SS_z}_{\lambda_1\lambda_2}$ constructs a state of definite spin
($S,S_z$) out of light-front helicity ($\lambda_1,\lambda_2$)
eigenstates. Explicitly,
 \be
 R^{SS_z}_{\lambda_1 \lambda_2}(x,k_\bot)
 =\sum_{s_1,s_2} \langle \lambda_1|
  {\cal R}_M^\dagger(1-x,k_\bot, m_1)|s_1\rangle
 \langle \lambda_2|{\cal R}_M^\dagger(x,-k_\bot, m_2)
 |s_2\rangle \langle \frac{1}{2}\,\frac{1}{2};s_1
 s_2|\frac{1}{2}\frac{1}{2};SS_z\rangle,
 \en
where $|s_i\rangle$ are the usual Pauli spinors, and ${\cal R}_M$ is
the Melosh transformation operator~\cite{Jaus1,Jaus91}:
 \be
 \langle s|{\cal R}_M (x_i,k_\bot,m_i)|\lambda\rangle
 &=&\frac{m_i+x_i M_0
 +i\vec \sigma_{s\lambda}\cdot\vec k_\bot \times
 \vec n}{\sqrt{(m_i+x_i M_0)^2 + k^2_\bot}},
 \en
with $x_1=1-x$, $x_2=x$, and 
$\vec n =(0,0,1)$ as a unit vector in the $\hat {z}$ direction. In
addition,
 \be
 M_0^2&=&(e_1+e_2)^2=\frac{m_1^2+k^2_\bot}{1-x}+\frac{m_2^2+k^2_\bot}{x},
 \\
 e_i&=&\sqrt{m^2_i+k^2_\perp+k^2_z}.\non
 \en
where $k_z$ is the relative momentum in $\hat{z}$ direction and can
be written as
 \be \label{eq:Mpz}
  k_z=\frac{x M_0}{2}-\frac{m^2_2+k^2_\perp}{2 x M_0}.
 \en
$M_0$ is the invariant mass of $q\bar q$ and generally different
from the mass $M$ of the meson which satisfies $M^2=P^2$. This is
due to the fact that the meson, quark and antiquark cannot be
simultaneously onshell. We normalized the meson state as
 \be
 \langle M(P',S',S'_z)|M(P,S,S_z)\rangle = 2(2\pi)^3 P^+
 \delta^3(\tilde P'- \tilde P)\delta_{S'S}\delta_{S'_z S_z}~,
 \label{wavenor}
 \en
which led to Eq. (\ref{normal}). 

In practice, it is more convenient to use the covariant form of
$R^{SS_z}_{\lambda_1\lambda_2}$ \cite{Jaus1,Jaus91,CCH2,cheung}:
 \be
 R^{SS_z}_{\lambda_1\lambda_2}(x,k_\bot)
 =\frac{\sqrt{p_1^+ p_2^+}}{\sqrt2~{\widetilde M_0}(M_0+m_1+m_2)}
 \bar u(p_1,\lambda_1)(\not\!\!\bar P+M_0)\Gamma
 v(p_2,\lambda_2), \label{covariantp}
 \en
where
 \be
 &&\widetilde M_0\equiv\sqrt{M_0^2-(m_1-m_2)^2},\qquad\quad \bar
 P\equiv p_1+p_2,\non \\
 &&\bar u(p,\lambda) u(p,\lambda')=\frac{2
 m}{p^+}\delta_{\lambda,\lambda'},\qquad\quad \sum_\lambda u(p,\lambda)
 \bar u(p,\lambda)=\frac{\not\!p +m}{p^+},\non \\
 &&\bar v(p,\lambda) v(p,\lambda')=-\frac{2
 m}{p^+}\delta_{\lambda,\lambda'},\qquad\quad \sum_\lambda v(p,\lambda)
 \bar v(p,\lambda)=\frac{\not\!p -m}{p^+}.
 \en
For the pseudoscalar meson, we have $\Gamma=\gamma_5$,
Eq. (\ref{covariantp})
can then be further reduced by the applications of equations of
motion on spinors \cite{CCH2}:
 \be
 R^{SS_z}_{\lambda_1\lambda_2}(x,k_\bot)
 =\frac{\sqrt{p_1^+ p_2^+}}{\sqrt2~{\widetilde M_0}}
 \bar u(p_1,\lambda_1)\gamma_5 v(p_2,\lambda_2). \label{covariantfurther}
 \en
Next, we derive the formulas of the decay constant and the mean
square radius for the pseudoscalar meson. The former is the main
subject of this work, and the latter is used to fix some parameters.

\subsection{Formulas for decay constant and mean square radius}
The decay constants of pseudoscalar mesons $P(q_1\bar{q}_2)$ are
defined in Eq. (\ref{definefp}). The matrix element can be
calculated using the formulism in the last subsection:
 \be
 \langle 0|\bar{q}_2\gamma_\mu\gamma_5q_1|P(P)\rangle &=& \int \{d^3p_1\}\{d^3p_2\}
 2(2\pi)^3\delta^3(\tilde P-\tilde p_1-\tilde
 p_2)\phi_P(x,k_\perp)R^{00}_{\lambda_1\lambda_2}
 (x,k_\perp)   \non \\
 && \times\,\langle 0|\bar{q}_2\gamma_\mu\gamma_5q_1|q_1\bar{q}_2\rangle.
 \en
Since $\widetilde{M}_0\sqrt{x(1-x)}=\sqrt{A^2+k^2_\perp}$, the decay
constant can be extracted as:
 \be
 f_P=\,2\sqrt{2N_c}\int \{dx\}\frac{A}{\sqrt{A^2+k_\perp^2}}
 \phi_P(x, k_\perp) \label{fp}
 \en
where $\{dx\}=\frac{dxd^2k_\bot}{16\pi^3}$ and $A=m_1x+m_2(1-x)$.

Next, the EM form factor of a meson $P$ is determined by the
scattering of one virtual photon and one meson. It describes the
deviation from the pointlike structure of the meson, and is a
function of $Q^2$. Here, we considered the momentum of the virtual
photon in a spacelike region, so it was always possible to orient
the axes in such a manner that $Q^+ = (P'-P)^+=0$. Thus, the EM form
factor was determined by the matrix element:
 \be
 \la P(P')|J^+|P(P)\ra= e~F_P(Q^2)
 (P+P')^+, \label{FPdef}
 \en
where $J^\mu=\bar q e_q e\gamma^\mu q$, $e_q$ is the charge of quark
$q$ in $e$ unit, and $Q^2=(P'-P)^2< 0$. With the light-front
framework, $F_P$ can be extracted by Eq. (\ref{FPdef}):
 \be
 F_P(Q^2)&=&e_{q_1}\int \{dx\} \frac{A^2+k_\perp\cdot k'_\perp}
 {\sqrt{A^2+k^2_\perp}\sqrt{A^2+k'^2_\perp}}
 \phi_P(x,k_\perp)\phi_{P'}(x,k'_\perp)\non \\
 &+&e_{\bar {q_2}}\int \{dx\} \frac{A^2+k_\perp\cdot k''_\perp}
 {\sqrt{A^2+k^2_\perp}\sqrt{A^2+k''^2_\perp}}
 \phi_P(x,k_\perp)\phi_{P'}(x,k''_\perp),
\label{FPgeneral}
 \en
where $k'_\perp=k_\perp+x Q_\perp$, $k''_\perp=k_\perp-(1-x)
Q_\perp$. 
For applying this to Eq. (\ref{MSR}), it is convenient to consider
the term $\widetilde {\phi}_{P} \equiv {\phi_{P}(x,k_\perp)/{\sqrt{
A^2+k^2_\perp}}}$ and take the Taylor expansion around $k^2_\perp$
 \be
 \widetilde {\phi}_{P'}(k'^2_\perp)=\widetilde
{\phi}_{P'}(k^2_\perp)+{d\widetilde
{\phi}_{P'}\over{dk^2_\perp}}\Bigg|_{Q_\perp=0}(k'^2_\perp-k^2_\perp)+{d^2\widetilde
{\phi}_{P'}\over{2(dk^2_\perp)^2}}\Bigg|_{Q_\perp=0}(k'^2_\perp-k^2_\perp)^2+.....
 \en
Then, by using the identity
 \be
 \int d^2k_\perp~(k_\perp \cdot A_\perp)(k_\perp \cdot B_\perp)={1\over{2}}\int
d^2k_\perp~k^2_\perp~A_\perp\cdot B_\perp, \label{QQQ2}
 \en
we can rewrite (\ref{FPgeneral}) to
 \be
 F_P(Q^2)&=&(e_{q_1}+e_{\bar q_2})\non \\
 &-&Q^2 \int \{dx\}\,\phi^2_P(x,k_\perp)\left[x^2 e_{q_1}+(1-x)^2e_{\bar q_2}\right]
 \Bigg( \Theta_{P}\frac{A^2+2k^2_\perp}{A^2+k^2_\perp}
 +\widetilde{\Theta}_{P}k^2_\perp\Bigg)\non \\
 &+&{\cal O}(Q^4), \label{FPQQ}
 \en
where
 \be
 \Theta_{M}={1\over{\widetilde {\phi}_{M}}}\Bigg({d\widetilde
 {\phi}_{M}\over{dk^2_\perp}}\Bigg),~~\widetilde{\Theta}_{M}={1\over{\widetilde
 {\phi}_{M}}}\Bigg({d^2\widetilde
 {\phi}_{M}\over{(dk^2_\perp)^2}}\Bigg).
 \en
It should be realized that the size and the density of a hadron
depend on the probe. For an EM probe, it is the electric charge
radius $\langle r^2 \rangle^{1/2}$ that is obtained. In the
experimental view, $\langle r^2_P \rangle$ cannot be measured
directly and is obtained by fitting the slope of $F_P(Q^2)$ at
$Q^2=0$, i.e.,
 \be
 \la r^2_P \ra= 6{dF_P(Q^2)\over{dQ^2}}\Bigg|_{Q^2=0}.\label{MSR}
 \en
Here the mean square radius is easily obtained: 
 \be
 \la r^2_P \ra &=& \la r^2_{q_1} \ra+\la r^2_{{\bar q}_2} \ra \non \\
 &=&e_{q_1}\Big\{-6\int\{dx\}x^2\widetilde {\phi}_{P}
 \Bigg[(A^2+2k^2_\perp)\frac{d}{dk^2_\perp}+(A^2+k^2_\perp)
 k^2_\perp\Bigg(\frac{d}{dk^2_\perp}\Bigg)^2\Bigg]\widetilde {\phi}_{P}\Big\}\non \\
 &+&e_{\bar{q}_2}\Bigg\{-6\int\{dx\}(1-x)^2\widetilde
 {\phi}_{P}\Bigg[(A^2+2k^2_\perp)\frac{d}{dk^2_\perp}+(
 A^2+k^2_\perp)k^2_\perp\Bigg(\frac{d}{dk^2_\perp}\Bigg)^2\Bigg]\widetilde
 {\phi}_{P}\Bigg\}.\label{MSRP}
 \en
It is worth mentioning that, first, the static property $F_P(0)=e_P$
is quite easily checked in Eq. (\ref{FPQQ}). Second, from Eq.
(\ref{MSRP}), we find that the mean square radius is related to the
first and second longitudinal momentum square derivatives of
$\widetilde {\phi}$ which contain the Melosh transformation effect.

If we take the heavy quark limit $m_1=m_Q\to \infty$, $m_Q (M_P)$ is
unimportant for the low energy properties of the meson state, so it
is more natural to use velocity $v$ instead of momentum variable
$P$. The normalization of the meson state is rewritten as
\cite{CCHZ}
 \be\label{normQ}
 \langle P(v')|P(v)\rangle =2(2\pi)^3 v^+ \delta^3(\bar \Lambda v - \bar \Lambda
 v'),
 \en
where $\bar \Lambda = M_P -m_Q$ is the residual center mass of the
heavy meson and the meson states have a relation $|P(v)\rangle =
(M_P)^{-1/2} |P(P)\rangle$. In addition, since $x$ is the
longitudinal momentum fraction carried by the light antiquark, the
meson wave function should be sharply peaked near $x \sim
\Lambda_{QCD}/m_Q$. It is thus clear that $x \to 0$ and only terms
of the form $X\equiv x m_Q$ survive in the wave function as $m_Q \to
\infty$; that is, $X$ is independent of $m_Q$ in the heavy quark
limit. Therefore, the normalization of the wave function Eq.
(\ref{normal}) is rewritten as
 \be\label{normalQ}
 \int \frac{dX d^2 k_\bot}{2(2\pi)^3} |\varphi(X,k_\bot)|^2=1.
 \en
where $\varphi(X,k_\bot)=(m_Q)^{-1/2}\phi(x,k_\bot)$. Other
replacements are 
$A \to \widetilde{A} = X+m_{q_2}$ and $\widetilde{\phi}(x,k_\bot)
\to \widetilde{\varphi}(X,k_\bot)$. Thus we can rewrite 
(\ref{MSRP}) as
 \be
 \la r^2_{Qq_2} \ra &=& \la r^2_{Q} \ra+\la r^2_{{\bar q}_2} \ra \non \\
 &=&\frac{e_{Q}}{m_Q^2}\Big\{-6\int\{dX\}X^2\widetilde {\varphi}
 \Bigg[(\widetilde{A}^2+2k^2_\perp)\frac{d}{dk^2_\perp}+(\widetilde{A}^2+k^2_\perp)
 k^2_\perp\Bigg(\frac{d}{dk^2_\perp}\Bigg)^2\Bigg]\widetilde {\varphi}\Big\}\non \\
 &+&e_{\bar{q}_2}\Bigg\{-6\int\{dX\}\widetilde
 {\varphi}\Bigg[(\widetilde{A}^2+2k^2_\perp)\frac{d}{dk^2_\perp}+(
 \widetilde{A}^2+k^2_\perp)k^2_\perp\Bigg(\frac{d}{dk^2_\perp}\Bigg)^2\Bigg]\widetilde
 {\varphi}\Bigg\}.\label{MSRHQ}
 \en
The first term of Eq. (\ref{MSRHQ}) vanished when $m_Q \to \infty$.
This means that not only $\la r^2_{Qq_2} \ra$ is blind to the flavor
of $Q$, but also $\la r^2_{P} \ra$ is insensitive to $m_1$ for the
heavy meson. The former is the so-called flavor symmetry and the
latter will be proven in the numerical calculation.

\section{Numerical results and discussions}
In the nonrelativistic (NR) approximation, we substituted the
experimental data \cite{PDG08} to Eq. (\ref{noparameter}), and
obtained $f_{B_s}/f_B= (1.03\pm 0.02) f_{D_s}/f_D$. If one wanted to
evaluate $f_{B_s}/f_B$ and $f_{D_s}/f_D$ individually, then
$m_s=483$ MeV and $m_q=310$ MeV were the ``best-fit" values for the
pseudoscalar and vector light meson masses \cite{textbook1}. The
ratios were
 \be
 \frac{f_{D_s}}{f_D}\Bigg|_{NR}=1.226\pm 0.002,\quad \frac{f_{B_s}}
 {f_B}\Bigg|_{NR}=1.24\pm 0.02.
 \en
The former was a little smaller than the data \cite{PDG08,CLEO09}
$f_{D_s}/f_D|_{\rm exp}=1.27\pm 0.06$, and the latter was almost
larger than the other theoretical calculations (see Tables II and
III).

In the light-front framework, the momentum distribution amplitude
$\phi (x, k_\bot)$ or the wave function $\Psi(\vec r)$ in principle
is unknown unless all the coefficients $a_{n00}$ are obtained.
However, we may suppose $a^c_{100}=1$ or $a^g_{000}=1$, that is,
$\Psi(\vec r)=\Psi^c_{100}(\vec r)$ or $\Psi(\vec
r)=\Psi^g_{000}(\vec r)$ as a trial wave function to fit the
relevant data. The other $a_{n00}$'s will be subsumed if the
parameters appearing in the momentum distribution amplitude can not
satisfy all experimental results. In other word, the coefficients
$a_{nlm}$ are taken as another kind of parameter. Of course, based
on the principle of quantum mechanics, the physical meanings of
these new parameters are clear. Here we list the first
$\phi^{c(g)}_{n00}$:
 \be
 \phi^c_{100}(x,k_\perp)&=&8\bigg(\frac{2\pi}{\beta^3}
 \bigg)^{1/2}\sqrt{\frac{e_1 e_2}{x (1-x) M_0}}\left[\frac{\beta^2}{k_\perp^2
 +k_z^2+\beta^2}\right]^2,\label{Pow1s}\\
 \phi^g_{000}(x,k_\perp)&=&4\bigg(\frac{\pi}
 {\beta^2}\bigg)^{3/4}\sqrt{\frac{e_1 e_2}{x (1-x) M_0}}~{\rm
 exp}\bigg[-\frac{k_\perp^2+k_z^2}{2
 \beta^2}\bigg],\label{Gaussian1s}
 \en
and use the experimental data of $f_{\pi^+}=130.4 \pm 0.2$ MeV and
$\langle r^2_{\pi^+}\rangle^{1/2}=0.672\pm 0.008$ fm to fit the
parameters $m_q$ and $\beta_{qq}$. The results are $m_q=0.172
(0.251)$ GeV and $\beta_{qq}=0.555\mp0.011 (0.317\mp0.007)$ GeV for
$\phi^c_{100}
(\phi^g_{000})$. As for the strange quark mass, 
in Ref. \cite{CJ99} $m_s-m_u = 0.23$ GeV was obtained with some
interaction potentials, while in Ref. \cite{Jaus91} $m_s-m_u = 0.12$
GeV in the invariant meson mass scheme. So here we use the values
$m_s-m_u = 0.180\pm 0.050$ GeV and $f_{K^+}=155.5\pm 0.8$ MeV to fix
$\beta_{sq}$. The results are $\beta_{sq}=0.463^{-0.032}_{+0.054}
(0.354^{-0.009}_{+0.015})$ GeV for $\phi^c_{100} (\phi^g_{000})$.
The charge radius $\langle r^2_{K^+}\rangle^{1/2}$ and the mean
square radius $\langle r^2_{K^0}\rangle$ were calculated by these
parameters and were listed in Table I.
\begin{table}[ht!]
\caption{\label{tab:parameter} Charge radius $\langle
r^2_{K^+}\rangle^{1/2}$  and the mean square radius $\langle
r^2_{K^0}\rangle$ of the experiment, this work, and other
theoretical estimations. DS is Dyson-Schwinger equations; VMD$\chi$
is vector meson dominance plus an effective chiral theory; BS is
Bethe-Salpeter equation. }
\begin{tabular}{|c|c|c|c|c|c|c|c|}\hline
 & Experiment\cite{PDG08} & $\phi^c_{100}$ &  $\phi^g_{000}$ & DS\cite{DS} & VMD$\chi$\cite{VMD1}
 & pQCD\cite{WH} & BS \cite{BS} \\ \hline
 $\langle r^2_{K^+}\rangle^{1/2}$ (fm) & $0.560 \pm 0.031$ 
 &$0.710^{+0.033}_{-0.044}$ & $0.607^{+0.010}_{-0.012}$& $0.49$ & $0.616$ & $0.570$ & $0.62$\\
 $\langle r^2_{K^0}\rangle$ (fm$^2$) & $-0.077\pm 0.010$ 
 &$-0.121^{-0.036}_{+0.039}$ &$-0.072_{+0.019}^{-0.017}$ &$-0.020$ & $0.057$ & $-0.0736$ & $-0.085$\\\hline
\end{tabular}
\end{table}
We found that, on one hand, the value of $\langle
r^2_{K^+}\rangle^{1/2}$ ( $\langle r^2_{K^0}\rangle$) for
$\phi^c_{100}$ was too large (small) than that obtained in the
experiment. Then, the coefficients $a^c_{n00}$ for $n>1$ may be
taken as nonzero to correct the fitting of $\langle
r^2_{K^+}\rangle^{1/2}$ and $\langle r^2_{K^0}\rangle$. However, the
mean square radii of $\phi^c_{n00}$ is greater when $n$ is larger,
or $\langle r^2\rangle_{\phi^c_{100}} < \langle
r^2\rangle_{\phi^c_{200}} < \langle r^2\rangle_{\phi^c_{300}} < ...$
This means, for decreasing the value of $\langle
r^2_{K^+}\rangle^{1/2}$, the values of $a^c_{n00}$ must be
artificially arranged in order to cancel out the contributions of
$\phi^c_{n00}$ $(n\geq 2)$ mutually. It is too hard to achieve now.
On the other hand, the results for $\phi^g_{000}$ were consistent
with the experimental data. Therefore we only use the Gaussian-type
wave function, $\phi^g_{000}$, to the following calculation. By
combining Eq. (\ref{splitting}), the experimental data \cite{PDG08},
and the light quark mass in above, we obtained the ratios as
 \be
  \frac{\beta_{cs}}{\beta_{cq}}\Bigg|^g=1.20\pm 0.04, \qquad
 \frac{\beta_{bs}}{\beta_{bq}}\Bigg|^g=1.20\pm 0.05. \label{ratiosg}
 \en
Obviously, the $SU(3)$ symmetry breaking is the major contribution
to the ratios in Eq. (\ref{ratiosg}).

For the heavy quark masses, the quite different values were also
used in the model
calculations. For example, 
$m_c=1.38$ GeV and $m_b=4.76$ GeV which were fitted for the spectrum
of the $p$-wave charmonium and bottomonium states \cite{Hwang1p};
and $m_c=1.8$ GeV and $m_b=5.2$ GeV which were obtained from the
potential models and the variational principle \cite{CJBc}. Here the
values $m_c= 1.2$, $1.5$, $1.8$ GeV and $m_b=4.2$, $4.7$, $5.2$ GeV
were taken into account. By combining Eqs. (\ref{fp}),
(\ref{Gaussian1s}), (\ref{ratiosg}), and the quark masses
$m_{q(s)}=0.251(0.431)$ GeV, the dependences of $f_{D_s}/f_D$ on
$f_D$ with three different $m_c$'s and $f_{B_s}/f_B$ on $f_B$ with
three different $m_b$'s were shown in Fig.1 and Fig.2, respectively.
It was easily found that the ratios $f_{D_s}/f_D$ and $f_{B_s}/f_B$
were not only insensitive to the heavy quark masses $m_c$ and $m_b$,
but also insensitive to the decay constants $f_D$ and $f_B$,
respectively.
 \begin{figure}
 \includegraphics*[width=4in]{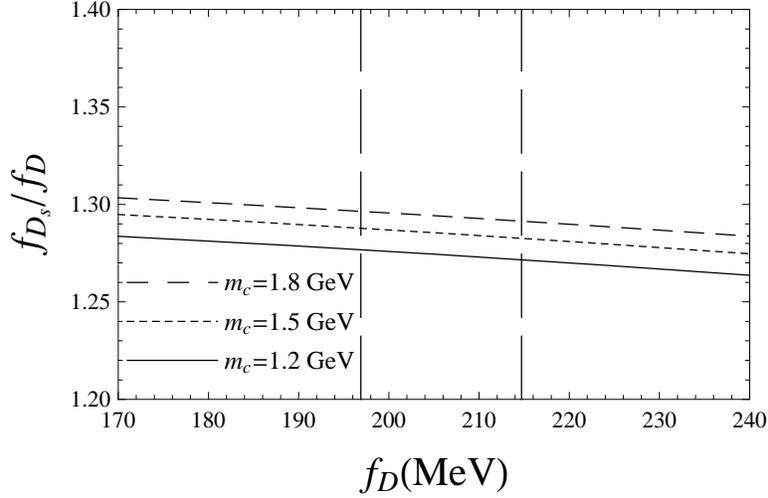}
 \caption{ Dependence of $f_{D_s}/f_D$ on $f_D$ with
$m_c = 1.2, 1.5, 1.8$ GeV. The left and right vertical dash lines
correspond to the lower and upper limits of the data $f_D=205.8\pm
8.9$ MeV, respectively. }
  \label{fig:DsD}
 \end{figure}
 \begin{figure}
 \includegraphics*[width=4in]{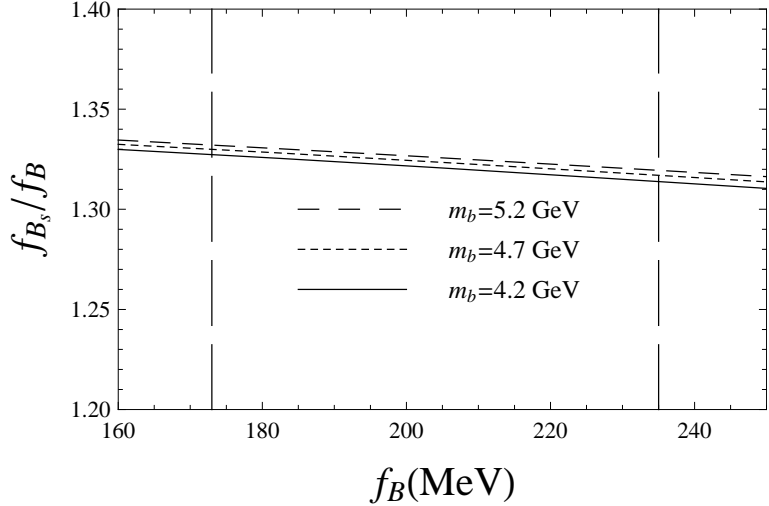}
 \caption{ Dependence of $f_{B_s}/f_B$ on $f_B$ with
$m_c = 4.2, 4.7, 5.2$ GeV. The left and right vertical dash lines
correspond to the lower and upper limits of the data $f_B=204\pm 31$
MeV, respectively. }
  \label{fig:BsB}
 \end{figure}

Recently, the CLEO collaboration updated their data about the
branching fraction for the purely leptonic decay $D^+ \to \mu^+ \nu$
and reported \cite{PDG08} $f^{\rm {exp}}_{D^+}=205.8 \pm 8.9$ MeV.
By using this value, we determined the ratio $f_{D_s}/f_D=1.29\pm
0.07$ and the decay constant $f_{D_s}=264.5\pm 17.5$ MeV with
$m_c=1.5$ GeV.
We found these results were consistent with the data \cite{CLEO09}:
$f^{\rm {exp}}_{D^+_s}=261.2 \pm 6.9$ MeV and $f^{\rm
{exp}}_{D^+_s}/f^{\rm {exp}}_{D^+}=1.27\pm 0.06$, which were the
average of the CLEO and Belle results (which included the radiative
corrections). 
In addition, our values were generally larger than
the other theoretical calculations. 
Table II compares the theoretical calculations with experimental
value.
\begin{table}[ht!]
\caption{\label{tab:compares} Theoretical calculations of the decay
constants $f_D$, $f_{D_s}$ (MeV), and the ratio $f_{D_s}/f_D$. QL is
quenched lattice calculations, BS is Bethe-Salpeter equation, Linear
and HO are the different potentials within LFQM. We have quoted only
the value with $m_c=1.5$ GeV in this work (LF). }
\begin{tabular}{|l|l|l|l|}\hline
  & $~~~~~f_D$  & $~~~~~f_{D_s}$ &  $~~~~f_{D_s}/f_D$  \\ \hline
 Experiment & $205.8\pm 8.9$ \cite{PDG08}  & $261.2\pm 6.9$ \cite{CLEO09} & $1.27\pm 0.06$
 \footnote{This value is obtained by combining $f_D=205.8\pm 8.9$ MeV \cite{PDG08} and
 $f_{D_s}=261.2\pm 6.9$ MeV \cite{CLEO09}.}\\
 This work (LF) & $\underline{205.8\pm 8.9}$  & $264.5\pm 17.5$ & $1.29\pm 0.07$
 \\
 This work (NR) & & & $1.226\pm 0.002$ \\
 Lattice (HPQCD+UKQCD) \cite{HU} & $208\pm 4$& $241\pm 3$ & $1.162\pm 0.009$ \\
 QL (QCDSF) \cite{QCDSF} & $206\pm 6\pm 3\pm 22$&$220\pm 6\pm5\pm11$ & $1.068\pm 0.018\pm 0.020$\\
 QL (Taiwan) \cite{CHIU} &$235\pm8\pm14$ &$266\pm10\pm18$ &$1.13\pm0.03\pm0.05$ \\
 Lattice (FNAL+MILC+HPQCD) \cite{MILC} & $201\pm 3\pm17$ & $249\pm3\pm16$ & $1.24\pm 0.01\pm 0.07$\\
 QL (UKQCD) \cite{UKQCD} &$210\pm10^{+17}_{-16}$ & $238\pm8^{+17}_{-14}$& $1.13\pm 0.02^{+0.04}_{-0.02}$ \\
 QL \cite{Bec} & $211\pm14^{+0}_{-12}$ & $231\pm12^{+6}_{-1}$ & $1.10\pm 0.02$ \\
 QCD Sum Rules \cite{BPS} &$177\pm21$ &$205\pm22$ & $1.16\pm0.01\pm0.02$\\
 QCD Sum Rules \cite{Nar} & $203\pm 23$ & $235\pm 24$ & $1.15\pm 0.04$ \\
 Field Correlators \cite{bada} & $210\pm 10$ & $260\pm 10$ & $1.24\pm 0.04$\\
 Potential Model\cite{EFG} & $234$ & $268$ & $1.15$ \\
 BS \cite{BS2} & $230\pm25$ & $248\pm27$ & $1.08\pm 0.01$ \\
 BS \cite{BS1} & $238$ & $241$ & $1.01$\\
 Linear\{HO\} \cite{Choi07} &$211\{194\}$ & $248\{233\}$ & $1.18\{1.20\}$\\\hline
\end{tabular}
\end{table}
For the bottom sector, the Belle \cite{Belle} and Babar
\cite{Babar1,Babar2} collaborations found evidence for $B^-\to
\tau^- \bar \nu$ decay which was not helicity suppressed. However,
the Belle and Babar values had $3.5$ and $2.6$ standard-deviation
significances, respectively; thus the average was provisional
\cite{RS}: ${\cal B} (B^-\to \tau^- \bar \nu)=(1.42 \pm 0.43)\times
10^{-4}$. We extracted the decay constant $f^{\rm exp}_B= 204\pm 31$
MeV. By using this value, the ratio $f_{B_s}/f_B=1.32\pm 0.08$ and
the decay constant $f_{B_s}=270.0\pm 42.8$ MeV with $m_b=4.7$ GeV
were obtained.
Table III compares the theoretical calculations with the
experimental value. Similar to the charm sector, our ratio
$f_{B_s}/f_B$ was almost larger than all other calculations.
\begin{table}[ht!]
\caption{\label{tab:comparesb} Theoretical calculations of the decay
constants $f_B$, $f_{B_s}$ (MeV), and the ratio $f_{B_s}/f_B$. Only
the value with $m_b=4.7$ GeV has been quoted in this work (LF). }
\begin{tabular}{|l|l|l|l|}\hline
  & $~~~~~f_B$  & $~~~~~f_{B_s}$ &  $~~~~f_{B_s}/f_B$  \\ \hline
 Experiment & $204\pm 31$ \footnote{This value is extracted by the branching ratio:
 ${\cal B} (B^-\to \tau^- \bar \nu)=(1.42 \pm 0.43)\times
10^{-4}$ \cite{RS}.}  &   & \\
 This work (LF) & $\underline{204\pm 31}$  & $270.0\pm 42.8$ & $1.32\pm 0.08$
 \\
 This work (NR) & & & $1.24\pm 0.02$ \\
 QL (QCDSF) \cite{QCDSF} & $190\pm 8\pm 23\pm 25$&$205\pm 7\pm26\pm17$ & $1.080\pm 0.028\pm 0.031$\\
 Lattice (HPQCD) \cite{HPQCD} & $216\pm 9\pm19\pm4\pm6$ & $259\pm32$ & $1.20\pm 0.03\pm 0.01$\\
 QL (UKQCD) \cite{UKQCD} &$177\pm17\pm22$ & $204\pm12^{+24}_{-23}$& $1.15\pm 0.02^{+0.04}_{-0.02}$ \\
 QL \cite{Bec} & $179\pm18^{+26}_{-9}$ & $204\pm16^{+28}_{-0}$ & $1.14\pm 0.03^{+0.00}_{-0.01}$ \\
 QCD Sum Rules \cite{BPS1} &$178\pm14$ &$200\pm14$ & $1.12\pm0.01\pm0.03$\\
 QCD Sum Rules \cite{Nar} & $203\pm 23$ & $236\pm 30$ & $1.16\pm 0.05$ \\
 QCD Sum Rules \cite{Jamin} & $210\pm 19$ & $244\pm 21$ & $1.16$ \\
 Field Correlators \cite{bada} & $182\pm 8$ & $216\pm 8$ & $1.19\pm 0.03$\\
 Potential Model \cite{EFG} & $189$& $218$ & $1.15$\\
 BS \cite{BS2} & $196\pm 29$ & $216\pm 32$ & $1.10\pm 0.01$\\
 BS \cite{BS1}& $193$ & $195$ & $1.01$\\
 Linear\{HO\} \cite{Choi07} &$189\{180\}$ & $234\{237\}$ & $1.24\{1.32\}$\\\hline
\end{tabular}
\end{table}
It is worth mentioning that the decay constants of both pseudoscalar
and vector heavy mesons have already been investigated by the author
of Ref. \cite{Choi07} with the analysis of magnetic dipole decays of
various heavy flavored mesons in the light-front quark model. The
parameters in Ref. \cite{Choi07} were constrained by the variational
principle for the QCD-motivated effective Hamiltonian. Roughly
speaking, our above results showed the flavor $SU(3)$ symmetry
breaking $m_s/m_q = 1.72\pm 0.20$ leads into the ratios
 \be
 \frac{f_{D_s}}{f_D}&=&1.29\pm 0.07, \non \\
 \frac{f_{B_s}}{f_B}&=&1.32\pm 0.08. \non
 \en

For the $B_c$ meson, however, both the decay constant and the
hyperfine splitting have not been measured yet. We considered the
ratio of hyperfine mass differences:
 \be
 \frac{\Delta
 M_{B_cB_c^*}}{\Delta M_{BB^*}}=\frac{m_q}{m_c}
 \left(\frac{\beta_{bc}}{\beta_{bq}}\right)^3.
 \en
Similar to the above cases, the ratio $f_{B_c}/f_B$ was insensitive
to the value of $m_b$ and sensitive to that of $m_c/m_q$. The
dependences of $f_{B_c}/f_B$ on $\Delta M_{B_cB_c^*}$ with $m_c =
1.2, 1.5, 1.8$ GeV and $m_b=4.7$ GeV were shown in Fig.3.
 \begin{figure}
 \includegraphics*[width=4in]{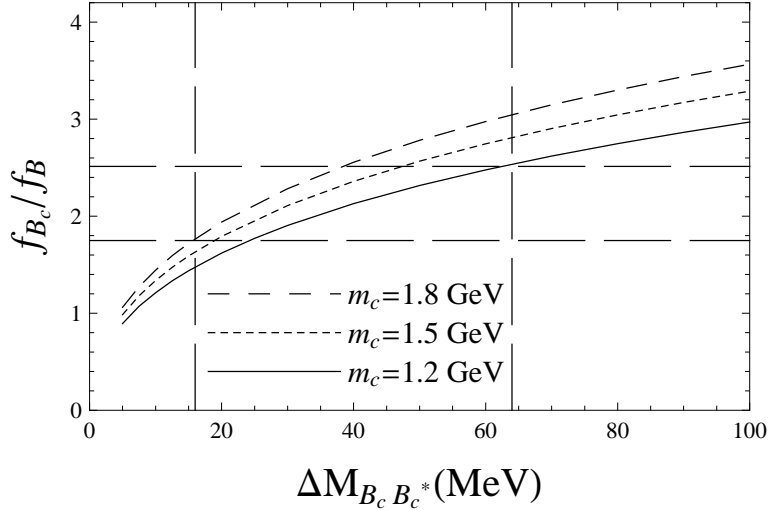}
 \caption{ Dependences of $f_{B_c}/f_B$ on $\Delta M_{B_cB_c^*}$ with
$m_c = 1.2, 1.5, 1.8$ GeV and $m_b=4.7$ GeV. The low and high
horizontal dash lines correspond to $f_{B_c}=360$ MeV and
$f_{B_c}=517$ MeV, respectively. The left and right vertical dash
lines correspond to $\Delta M_{B_cB_c^*}=16$ MeV and $\Delta
M_{B_cB_c^*}=64$ MeV, respectively. }
  \label{fig:phicb}
 \end{figure}
Some model predictions were made for $f_{B_c}$
\cite{CG,EQ,GKLT,LP,IKS,EFG,CJBc}, and the range of these values was
$f_{B_c}=360\sim 517$ MeV or $f_{B_c}/f_B = 1.76\sim 2.53$. As shown
in Fig.3, this range corresponded to $\Delta M_{B_cB_c^*}=16\sim 64$
MeV. We found that this result was consistent with a calculation
using the nonrelativistic renormalization group \cite{PPSS} $\Delta
M_{B_cB_c^*}=48\pm 15^{+14}_{-11}$ MeV.

Besides, the mean square radii of the heavy meson are calculated by
the above parameters and Eq. (\ref{MSRP}). The dependences of
$\langle r^2_{D^+,D^0,D_s}\rangle$ on $m_c$ and $\langle
r^2_{B^+,B^0,B_s}\rangle$ on $m_b$ were shown in Fig.4 and Fig.5,
respectively. It was easily found that, as mentioned at the end of
sec. III, the mean square radii $\langle r^2_{D^+,D^0,D_s}\rangle$
and $\langle r^2_{B^+,B^0,B_s}\rangle$ were insensitive to the heavy
quark masses $m_c$ and $m_b$, respectively.
 \begin{figure}
 \includegraphics*[width=4in]{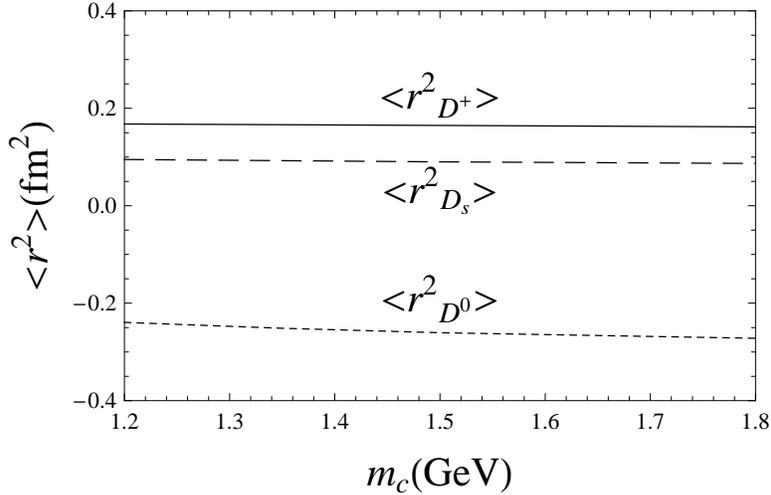}
 \caption{ Dependences of $\langle
r^2_{D^+,D^0,D_s}\rangle$ on $m_c = 1.2 \sim 1.8$ GeV. }
  \label{fig:RDsD}
 \end{figure}
 \begin{figure}
 \includegraphics*[width=4in]{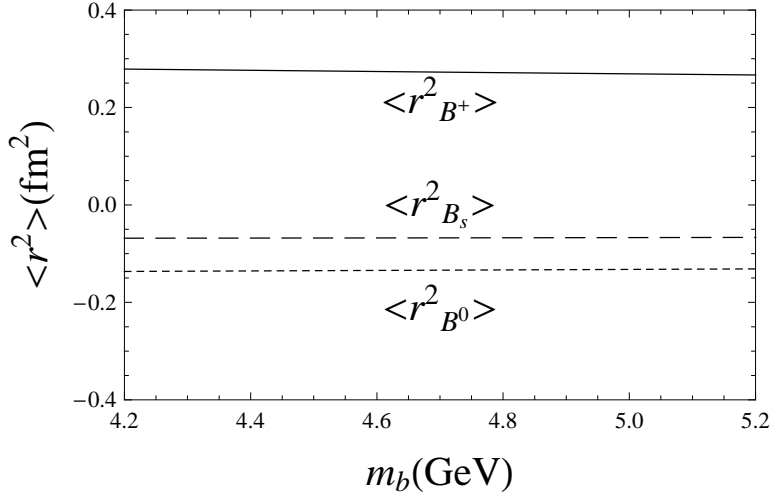}
 \caption{ Dependences of $\langle r^2_{B^+,B^0,B_s}\rangle$ on
$m_c = 4.2 \sim 5.2$ GeV. }
  \label{fig:RBsB}
 \end{figure}
We used $m_c=1.5$ GeV and $m_b=4.7$ GeV to estimate the mean square
radii of the heavy meson and the results are listed in Table IV.
\begin{table}[ht!]
\caption{\label{tab:MSR} Mean square radius $\langle r^2_P\rangle$
(fm$^2$) 
of this work and the other theoretical calculations.}
\begin{tabular}{|c|c|c|c|}\hline
 & This work & Lattice \cite{latticeQ} & VMD \\\hline
 $D^+$ & $0.165^{-0.010}_{+0.011}$  & & \\
 $D^0$ &   $-0.261^{+0.018}_{-0.019}$ & & \\
 $D_s^+$ &  $0.0902^{-0.011}_{+0.014}$ & & \\
 $B^+$ &  $0.273^{-0.043}_{+0.059}$  & $0.334\pm 0.003$& $0.393$
 \footnote{This value is obtained by $\langle r^2\rangle_{\rm VMD}=6/M^2_\rho$.}\\
 $B^0$ &   $-0.134^{+0.022}_{-0.029}$ & & \\
 $B_s^0$ &  $-0.0676^{+0.0141}_{-0.0189}$ & & \\
 $B_c^+$ &  $0.0277\sim 0.0451$ \footnote{This value is obtained by $f_{B_c}=360\sim 517$ MeV.}& & \\ \hline
\end{tabular}
\end{table}
It is interesting to note that the value $\langle r^2_{B^+} \rangle$
is slightly lower but comparable to what one would obtain from the
lattice calculation of Ref. \cite{latticeQ}, and it is considerably
smaller than the results obtained by applying the simple vector
meson dominance. The $SU(3)$ symmetry breaking in $\langle
r^2_{P}\rangle^{1/2}$, which mainly come from the mass difference
$m_s-m_q = 180\pm 50$ MeV, are obtained as
 \be
 \sqrt{\frac{\langle r^2_{D_s^+}\rangle}{\langle r^2_{D^+}\rangle}} &=&0.740^{-0.041}_{+0.050}, \non \\
 \sqrt{\frac{\langle r^2_{B_s^0}\rangle}{\langle r^2_{B^0}\rangle}} &=&0.711^{-0.049}_{+0.058}.  \non
\en
The radius ratio of $B_c^+$ and $B^+$ are also obtained as
 \be
 \sqrt{\frac{\langle r^2_{B_c^+}\rangle}{\langle r^2_{B^+}\rangle}}
 &=&0.407^{+0.037}_{-0.038}\sim 0.319^{+0.029}_{-0.030}, \non
 \en
which corresponds to the range $f_{B_c}=360\sim517$ MeV.
\section{Conclusions}
In this study, we discussed the ratios of decay constants and mean
square radii for pseudoscalar heavy mesons. By considering the
hyperfine interaction inside the meson, we found that the ratio of
light quark masses $m_s/m_q$ was the important factor for
determining the ratio of the decay constants. First, in the
nonrelativistic approximation, we obtained the relation $f_{B_s}/f_B
= (1.05\pm 0.02) f_{D_s}/f_D$ which did not use any parameters.
These two ratios were individually evaluated by including the
``best-fit" light quark masses $m_s/m_q=483/310=1.558$ and the
values $f_{D_s}/f_D = 1.226\pm 0.002$ and $f_{B_s}/f_B = 1.24\pm
0.02$ were obtained. Second, in the light-front framework, we
utilized the mass difference of light quark masses $m_s-m_q=180\pm
50$ MeV and the fittings of the decay constants for light mesons to
compare the mean square radii of $K^{+,0}$ mesons in the power-law
and Gaussian momentum distribution amplitudes. The latter was
consistent with the data and it extracted the light quark masses
ratio $m_s/m_q=1.72\pm0.20$. This mass ratio led to $f_{D_s}/f_D =
1.29\pm 0.07$ and $f_{B_s}/f_B = 1.32\pm 0.08$. The former was in
agreement with the experimental data and the latter was almost
larger than all other theoretical calculations. Both these ratios
were not only insensitive to the heavy quark masses $m_{c,b}$, but
also insensitive to the decay constants $f_{D,B}$. Similar to the
above, the ratio $f_{B_c}/f_B$ was mainly determined by the mass
ratio $m_c/m_q$ and the mass splitting $\Delta M_{B_cB^*_c}$. The
dependences of $f_{B_c}/f_B$ on $\Delta M_{B_cB^*_c}$ with the
varied charm quark masses have been shown. We found that
$f_{B_c}/f_B = 1.76\sim 2.53$ corresponded to $\Delta
M_{B_cB_c^*}=16\sim 64$ MeV. In addition, the mean square radii of
heavy meson were estimated. We found the mean square radii $\langle
r^2_{D^+,D^0,D_s}\rangle$ and $\langle r^2_{B^+,B^0,B_s}\rangle$
were insensitive to the heavy quark masses $m_c$ and $m_b$,
respectively, which was consistent with the behavior when the heavy
quark limit was taken. Our $\langle r^2_{B^+} \rangle$ was slightly
lower but comparable to that of lattice calculation, and was
considerably smaller than that of vector meson dominance (VMD). The
light quark mass ratio and the range of $f_{B_c}$ given above also
led the radius ratios $\sqrt{\langle r^2_{D_s^+}\rangle/\langle
r^2_{D^+}\rangle} =0.740^{-0.041}_{+0.050}$, $\sqrt{\langle
r^2_{B_s^0}\rangle/\langle r^2_{B^0}\rangle}
=0.711^{-0.049}_{+0.058}$, and $\sqrt{\langle
r^2_{B_c^+}\rangle/\langle r^2_{B^+}\rangle}
 =0.407^{+0.037}_{-0.038}\sim 0.319^{+0.029}_{-0.030}$, respectively.

{\bf Acknowledgements}\\
The author would like to thank Shu-Yin Wang for her helpful
discussion. This work was supported in part by the National Science
Council of the Republic of China under Grant No
NSC-96-2112-M-017-002-MY3.


\end{document}